\documentclass[prd,showkeys,nofootinbib]{revtex4-2}
\usepackage{amssymb, amsmath,amsthm,graphicx,hyperref,subfigure,braket,microtype}

\begin{document}
\title{Phenomenological inflationary model in Supersymmetric Quantum Cosmology}
\author{N.E. Mart\'inez-Pérez}
\email{nephtalieliceo@hotmail.com}
	
\author{C. Ramírez}
\email{cramirez@fcfm.buap.mx}
\affiliation{Facultad de Ciencias
F\'{\i}sico Matem\'aticas, Benem\'erita Universidad Aut\'onoma de Puebla. Puebla 72000, M\'exico.}

\author{V. Vázquez-Báez}
\email{manuel.vazquez@correo.buap.mx}
\affiliation{Facultad de Ingeniería, Benemérita Universidad Autónoma de Puebla, \mbox{Puebla 72000, Mexico;}}
\date{\today}
	
\begin{abstract}
We consider the effective evolution of a phenomenological model from  FLRW supersymmetric quantum cosmology with a scalar field. The scalar field acts as a clock and inflaton. We examine a family of simple superpotentials that produce an inflation whose virtual effect on inhomogeneous fluctuations shows very good agreement with PLANCK observational evidence for the tensor-to-scalar ratio and the scalar spectral index.
\end{abstract}

\maketitle

\section{Introduction}
\label{intro}
It is widely accepted that the observable universe originates from an early homogeneous phase beginning presumably around the Planck scale, just after a less understood phase of quantum spacetime. This homogeneous phase has a classical description by an effective FLRW model with scalar matter. The subsequent inhomogeneity is attributed to quantum fluctuations of space that induce matter inhomogeneities that grow, and loss coherence due to inflation. Thus, a quantum treatment seems to be natural for the homogeneous phase. 

Standard quantum cosmology is based on the canonical quantization of general relativity~\citep{Hartle}. Following Dirac, the Hamiltonian constraint, generator of time reparametrizations, is implemented as a time-independent Schr\"odinger equation, the Wheeler--DeWitt (WDW) equation. Additionally, the solution to the WDW equation, i.e., the wave function of the universe, must be supplemented with a prescription to compute probabilities considering the singular character of the universe. Further, as the Hamiltonian operator annihilates the wave function, this is a timeless theory. However, the universe exhibits a marked one-way evolution (arrow of time), and internal observers have clocks \citep{tiempo,Kuchar,Isham,Anderson}.

In recent years, there have been several proposals in standard quantum cosmology that mainly relate time to the scale factor, essentially by a gauge fixing. These proposals involve mostly approximations of the semiclassical type, as well as Born--Openheimer with a weakly coupled matter sector and a large-scale factor. In~\citep{barvisnky}, a general gauge fixing was analyzed, which was applied in~\citep{barvisnkyk} to several minisuperspace models with a classical `extrinsic' time. In~\citep{venturi98}, a Born--Oppenheimer approximation was used to separate the scale factor from a weakly coupled matter sector. This allowed for defining a time as in the WKB approach, proportional to the square of the scale factor, and parametrize matter evolution with an effective Hamiltonian; see also~\citep{moniz1,moniz2}. Further, in~\citep{venturi98} and a subsequent series of works, increasingly complex settings were  analyzed, an inflaton with a mass term in~\citep{venturi98}, additional generic matter in~\citep{venturi03}, Mukhanov--Sasaki scalar perturbations for a de Sitter evolution in~\citep{venturi13}, and tensor perturbations in a general slow-roll inflationary setting in~\citep{venturi14}. The parameter values of the last work were restricted in~\citep{venturi15} by comparison with observational data, and in~\citep{venturi16}, they  were considered  effects on the spectra of primordial perturbations. In~\citep{venturi18} the consequences of the interference of the wave functions before and after a bounce were analyzed. For supersymmetric quantum cosmology, in~\citep{moniz1,moniz2,obregon2}, in the semiclassical approach, the effect of supersymmetry was explored for the solution. 

The purely bosonic Wheeler--DeWitt equation is a second-order partial differential equation (PDE) in (mini)superspace. To single out a solution that corresponds to the wave function of the universe, one must impose suitable boundary conditions.  Defining the appropriate boundary conditions that give rise to the universe with which we are familiar  constitutes a well-known fundamental problem in quantum cosmology. There is also the possibility  that a more complete theory introduces additional restrictions that uniquely determine a quantum state. An example of this is quantum supersymmetric cosmology; see, e.g.,~\citep{moniz1,moniz2}. In this case, the most general state has multiple components, and the Hamiltonian constraint amounts to a system of coupled second-order PDEs that is equivalent to a system of first-order PDEs, the supersymmetric constraints. In~\citep{previo:2016}, we worked out a supersymmetric model leading to an analytic solution to the WDW equation. In the present work, with a slightly simpler WDW equation by a different operator ordering, we consider the time choice of~\citep{previo:2016} by analyzing the wave function. The general expression of this wave function allows for the identification of a curve of most probable values in configuration space (superspace), parametrized by the scalar field, and suggests to choose it as time. Such a type of choice has been known for a long time~\citep{banks}. Here, we analyze in detail the wave function regarding the choice  in~\citep{previo:2016}, of an effective time-dependent wave function with a probability density that corresponds to the conditional probability of measuring a value of the scale factor for a given value of the scalar field. Thus, the mean values computed with this wave function were time-dependent, in particular for the scale factor, which gives a trajectory that follows closely the previously mentioned curve of most probable values. Further, we consider the quantum cosmology of this approach, and explore a family of inflationary solutions.

We consider supersymmetric models because, even if at LHC energies, supersymmetry has not been found, it might be broken at higher energies, and it continues to be part of important candidates of ultraviolet completions for quantum gravity, supergravity, and string theory. Supersymmetry nontrivially relates   fermions and bosons; in a supersymmetric quantum field theory, fermionic and bosonic divergences are cancelled~\citep{wess}. 
Thus, supersymmetric quantum cosmology~\citep{obregon0,eath,moniz-1,moniz-2} is a relevant option for the study of quantum cosmology. Supersymmetry can be formulated by the extension of spacetime translations to translations in a Grassmann-extended spacetime, which includes fermionic coordinates, called superspace\footnote{Should not be confused with the superspace of geometrodynamics.}. The fields on this supersymmetry superspace are called superfields, and supergravity can be formulated as a general relativity theory on a supermanifold~\citep{wess,cupa}. There are several formulations for supersymmetric extensions of homogeneous cosmological models~\citep{eath,moniz-1}. One class of such formulations comes from a dimensional reduction of four- or higher-dimensional supergravity theories by considering homogeneous fields depending only on time and integrating the space coordinates~\citep{ryan}. The other class is obtained by supersymmetric extensions of homogeneous models, invariant under general reparametrizations of time~\citep{graham,graham1,graham2}, or invariant under general reparametrizations on a superspace, with anticommutative coordinates besides time~\citep{tkachwf,previo}. In~\citep{previo:2016,previo}, we followed~\citep{cupa}, where one of us gave a geometric Lorentz covariant superfield model building tool for supergravity that can be straightforwardly applied to any number of spacetime dimensions. In particular, a homogeneous formulation could be straightforwardly traced back to four-dimensional supergravity.

As usual in theories with fermionic degrees of freedom, their conjugate momenta are eliminated by solving a number of second-class constraints, leaving an algebra of Dirac brackets. The corresponding quantum fermionic operators can be represented in several ways, e.g., by a matrix representation~\citep{obregon0,obregon1},  representing half the fermionic degrees of freedom by either differential operators~\citep{eath2}, or by creation operators acting on a vacuum state~\citep{graham2,previo:2016}. The wave function is spinorial in the first case, or a finite expansion of possible states in the second case. In any case, the Wheeler--DeWitt equation is equivalent to a system of first-order partial differential equations. In many cases, they have exact solutions~\citep{tkachwf,obregon1,previo}, and the integration constants can be assimilated in the normalization. Therefore no initial conditions are required, although the state  depends on the model. For supersymmetric extensions of higher-order theories, such as $f(R)$ theories, the differential equations might not be first-order~\citep{nephtali}.

In Section \ref{FLRW}, we review the superfield formulation of supersymmetric cosmology. In Section \ref{quantization}, we review the quantization of the models from~\citep{previo:2016};  supersymmetric Wheeler--DeWitt equations have an analytic solution that depends on the scale factor and the superpotential. In Section \ref{time}, we discuss the problem of time. The identification of high-probability paths in configuration space to which mean trajectories correspond leads to the identification of the scalar field as time. Thus, following~\citep{previo:2016}, a time-dependent effective wave function can be given. This effective wave function allows for computing mean values of the scale factor that give a time evolution. This scale factor is inversely proportional to the cubic root of the superpotential, and we obtain inflationary behavior for a family of superpotentials, as shown in Section \ref{inflation}. These superpotentials depend on three parameters, namely, $\mu$, $\lambda$, and $p$; the first  determines the time scale, the second  the number of $e$-folds, and the last is a power that modifies the initial conditions of the scale factor. After fixing  parameters $\mu$ and $\lambda$, we compute the tensor/scalar ratio and the scalar spectral index  with remarkably good agreement with the observational bounds of PLANCK observatory without having to adjust parameters for this purpose. Lastly, in Section \ref{discussion}, we  present a short discussion with some remarks and future work perspectives. In an appendix, we give the effective wave function and the scale factor for $k=1$.

\section{Supersymmetric FLRW Model with a Scalar Field}
\label{FLRW}

The large-scale observable universe has been modeled in general relativity by the FLRW metric with scalar fields. This is  quite a general setting that could follow from a fundamental theory, and can account for inflation, primordial matter generation and structure formation, and dark energy. We consider the most studied model, the simplest, with a single minimally coupled scalar field $\frac{1}{2\kappa^2}\int\sqrt{-g}Rd^4x+\int\sqrt{-g}[\frac{1}{2}\partial^\mu\phi\partial_\mu-V(\phi)]d^4x$, where $\kappa^2=\frac{8\pi G}{c^4}$. For the FLRW metric, it reduces to  well-known form
\begin{equation}
	I=\frac{1}{\kappa^{2}}\int\left\{  -\frac{3}{c^2}N^{-1}a\dot{a}^{2}+3Nka-Na^3\Lambda+\kappa^{2}a^{3}\left[\frac{1}{2c^2}N^{-1}\dot{\phi}^{2}-NV\left(\phi\right)\right]\right\}dt.
	\label{actioncosmo}
\end{equation}

This Lagrangian is invariant under general time reparametrizations. From this action, follow the Friedmann equations and the conservation equation for a perfect fluid described by  scalar field $\phi(t)$, i.e., in natural units and comoving gauge,
$\frac{\dot{a}^{2}}{a^{2}}-\frac{\Lambda}{3}+\frac{k}{a^2}=\frac{\kappa^{2}}{3}\rho$, $\frac{2\ddot{a}}{a}+\frac{\dot{a}^{2}}{a^{2}}-\Lambda+\frac{k}{a^2}=-\kappa^{2}p$ and
$\dot{\rho}+\frac{3\dot{a}}{a}(p+\rho)=0$,
with $\rho=\frac{1}{2} \dot{\phi}^{2}+V(\phi)$ the energy
density, and $p=\frac{1}{2} \dot{\phi}^{2}-V(\phi) $ the pressure for the
perfect fluid $\phi(t)$.
The momenta are $\pi_a=-\frac{6}{c^2\kappa^{2}}N^{-1}a\dot{a}$ and $\pi_\phi=-\frac{1}{c^2}N^{-1}a^3\dot{\phi}$
The Hamiltonian is $H=NH_0$, where $H_0$ is the Hamiltonian  constraint, 
which generates time reparametrizations. 

\subsection{Supersymmetric Cosmology}
Supergravity, the supersymmetric version of gravity, can be formulated as general relativity in superspace, an extension of spacetime by anticommutative spinorial variables, in four dimensions, $x^m\to(x^m,\theta_\alpha,\bar{\theta}^{\dot{\alpha}})$, where $\theta$ are Weyl spinors and $\bar{\theta}$ their conjugates~\citep{wess}, $\alpha, \dot{\alpha}=1,2$. Supersymmetric field theory on superspace realizes supersymmetry algebra
$\left\{Q_\alpha,\bar{Q}_{\dot{\alpha}}\right\}=2i\sigma^m_{\alpha\dot{\alpha}}\partial_m$, which extends the Poincar\'e algebra. Thus, for homogeneous fields, the charges decompose into two copies 
\begin{equation}
	\left\{Q_\alpha,\bar{Q}_{\dot{\alpha}}\right\}=2i\sigma^0_{\alpha\dot{\alpha}}\partial_0=2i\delta_{\alpha\dot{\alpha}}\partial_0.
\end{equation}

Therefore, a minimal version of homogeneous supersymmetric field theory can be given by extending time by one complex anticommuting coordinate, which amounts to supersymmetric quantum mechanics. This theory can be obtained, in general, by dimensional reduction from higher-dimensional models to one (time) dimension~\citep{halpern}. Thus, supersymmetric cosmology can be obtained from one-dimensional supergravity~\cite{tkachwf}.

For the sake of clarity, here we shortly review the derivation of the supersymmetric Wheeler--DeWitt equation following~\citep{previo,previo:2016}. In these works, we formulated it as general relativity on supersymmetry--superspace,
$t\rightarrow z^{M}=(t,\Theta,\bar{\Theta})$, where $\Theta$ and $\bar{\Theta}$
are the anticommuting coordinates, the so-called “new” $\Theta$-variables~\citep{wess}. Hence, under $z^{M}\rightarrow z^{\prime M}=z^{M}+\zeta^{M}(z)$, the superfields,  see e.g.,~\cite{wess,cupa},
transform as $\delta_{\zeta}\Phi(z)=-\zeta^{M}(z)\partial_{M}\Phi(z)$,
and their covariant derivatives are $\nabla_{A}\Phi={\nabla_{A}^{\ M}(z)\partial_{M}}\Phi$. $\nabla_{A}^{\ M}(z)$ is the superspace vielbein,
whose superdeterminant gives the invariant superdensity
$\mathcal{E}=\mathrm{Sdet}{\nabla_{M}}^{A}$, $\delta_{\zeta}\mathcal{E}=(-1)^{m}\partial_{M}(\zeta^{M}\mathcal{E})$. 
For the supersymmetric extension of the FLRW metric, we have $\mathcal{E}=-N-\frac{i}{2}(\Theta\bar{\psi}+\bar{\Theta}\psi)$ \cite{previo}. In this formulation, to the scale factor and the scalar field correspond real scalar superfields~\citep{wess,tkachwf}. 
\begin{align}
	\mathcal{A}\left(  t,\Theta,\bar{\Theta}\right)  =a\left(  t\right)
	+\Theta{\lambda}\left(  t\right) -\bar{\Theta}\bar\lambda\left(  t\right)
	+  \Theta\bar{\Theta}B\left(t\right),\\
	\Phi\left(  t,\Theta,\bar{\Theta}\right)  =\phi\left(  t\right)  +\Theta{\eta}\left(  t\right)
	-\bar{\Theta}\bar\eta(t)+\Theta\bar{\Theta}G\left(  t\right). \label{superfields}
\end{align}

The supersymmetric extension of Action (\ref{actioncosmo}) for $k=0,1$ is
$I=I_G+I_{M}$ , where $I_{G}$ is the supergravity action, and $I_M$ is the matter term \cite{tkachwf,superfield,previo:2016}
\begin{align}
	I_{G}&=\frac{3}{\kappa^{2}}\int\mathcal{E}\left(  \mathcal{A}\nabla
	_{\bar{\Theta}}\mathcal{A}\nabla_{\Theta}\mathcal{A}-\sqrt{k}\mathcal{A}^{2}\right)  d\Theta d\bar{\Theta}dt, \label{susyraction}\\
	I_{M}&=\int\mathcal{E}\mathcal{A}^{3}\left[  -\frac
	{1}{2}\nabla_{\bar{\Theta}}\Phi\nabla_{\Theta}\Phi+W\left(  \Phi\right)
	\right]  d\Theta d\bar{\Theta}dt, \label{susymaction}
\end{align}
where $W$ is the superpotential.
\subsection{Component Formulation}
\label{componentform}
The component action follows from (\ref{susyraction}) and (\ref{susymaction}). After performing the Grassmann integrals, integrating out  auxiliary fields $B$ and $G$, and producing the redefinitions
$\lambda\rightarrow a^{1/2}\lambda$, $\bar{\lambda}\rightarrow a^{1/2}\bar{\lambda}$,
$\eta\rightarrow a^{3/2}\eta$, and $\bar{\eta}\rightarrow a^{3/2}\bar{\eta}$, the Lagrangian reads, see~\citep{previo:2016}
\begin{align*}
		L  =-\frac{3a\dot{a}^{2}}{c^2N\kappa^{2}}+\frac{a^{3}\dot{\phi}^{2}}{2c^2N}
		+\frac{3kNa}{\kappa^{2}}-3\sqrt{k}Na^{2} W+\frac{3N\kappa^{2}}{4}a^{3}W^{2}-\frac{1}{2}a^{3}N{W'}^{2}
		+\frac{3i}{c\kappa^{2}}\left(  \lambda\dot{\bar{\lambda}}+\bar{\lambda}\dot{\lambda}\right)
		-\frac{i}{2c}\left(  \eta\dot{\bar{\eta}}+\dot{\eta}\bar{\eta}\right) \\
		+\frac{3\sqrt{a}\dot{a}}{c\kappa^2N}\left(  \psi\lambda-\bar{\psi}\bar{\lambda}\right)
		-\frac{a^{3/2}\dot{\phi}}{2cN}\left(  \psi\eta-\bar{\psi}\bar{\eta}\right)
		+\frac{3i\dot{\phi}}{2c}\left(  \lambda\bar{\eta}+\bar{\lambda}\eta\right)
		+3N\left(\frac{\sqrt{k}}{\kappa^{2}a}-\frac{3}{2}W\right)\lambda\bar{\lambda}
		+N\left(-\frac{3\sqrt{k}}{2a}+\frac{3\kappa^{2}}{4}W-W''\right)\eta\bar{\eta} \\
		+3ia^{3/2}\left(\frac{\sqrt{k}}{\kappa^2a}-\frac{1}{2}W\right)\left(  \psi\lambda+\bar{\psi}\bar{\lambda}\right)
		-\frac{ia^{3/2}}{2}W'\left(  \psi\eta+\bar{\psi}\bar{\eta}\right)
		-\frac{3N W'}{2}\left(  \lambda\bar{\eta}-\bar{\lambda}\eta\right)
		-\frac{3}{2N\kappa^2}\psi\bar{\psi}\lambda\bar{\lambda}+\frac{1}{4N}\psi\bar{\psi}\eta\bar{\eta},
	\end{align*}
where $W\equiv W(\phi)$,  $W'\equiv \partial_\phi W(\phi)$, and $W''\equiv \partial_\phi^2 W(\phi)$.

The Hamiltonian is of the form $H=NH_{0}+\frac{1}{2}\psi S-\frac{1}{2}\bar{\psi}\bar{S}$, with the components of the one-dimensional supergravity multiplet $(N,\psi,\bar{\psi})$ as Lagrangean multipliers enforcing the Hamiltonian ($H_0\approx 0$) and supersymmetric ($S\approx 0$, $\bar{S} \approx 0$) constraints~\citep{previo:2016}. 

The basic nonvanishing Dirac brackets are $\{a,\pi_{a}\}=\{\phi,\pi_{\phi}\}=1$, $\{\lambda,\bar{\lambda}\}_{+}=\frac{c\kappa^2}{6}$, $\{\eta,\bar{\eta}\}_{+}=-c$. With these brackets, one can verify the following algebra of constraints
\begin{align}
	\{S,\bar S\}_{+}&=-2H_0,\label{ss}\\
	\{H_0,S\}&=\{H_0,\bar S\}=0. \label{hs}
\end{align}

The scalar potential in the Hamiltonian $H_0$ is~\citep{previo:2016}  
\begin{equation}
	V_S=\frac{3\sqrt{k}}{a}W-\frac{3\kappa^{2}}{4}W^{2} +\frac{1}{2}{W'}^{2}.\label{scalarpot}
\end{equation}

\textls[-15]{For $k=0$, the sign of the superpotential does not matter for the scalar~potential.}
\section{Quantization}
\label{quantization}
Homogeneous cosmology is a mechanical system; hence, it can be quantized with the formalism of ordinary quantum mechanics. There are, however, several well-known problems. On the one hand, the probabilistic character of measurements in quantum mechanics clashes with the uniqueness of the system as there is no ensemble of universes to perform a series of tests, in identical, observer shaped conditions~\citep{halliwell}. Nevertheless, observables such as the Hubble parameter can be determined by a set of observations. On the other hand, since the Hamiltonian vanishes, a time parameter cannot be introduced by means of the Schr\"odinger equation, or for the wave function or for the observables. Nonetheless, the Wheeler--DeWitt equation gives a time-independent Schr\"odinger equation with zero eigenvalue whose solution depends on the superspace variables, the minisuperspace in the homogeneous case, and gives the probability amplitude for the universe to be found in certain superspace configurations. The fact that the theory does not give a time evolution is the well-known consequence of invariance under time reparametrizations. Time is argued to be an internal property that can be determined by the choice of a clock~\citep{Kuchar}. On the other side, the observed universe is classical~\citep{halliwell}; hence, its description is given by mean values of the quantum operators. We further discuss the time problem in Section \ref{time}.
\subsection{Supersymmetric Wheeler--DeWitt Equations}
For the derivation of  supersymmetric Wheeler--DeWitt equations, we follow~\citep{previo:2016}, but with a different ordering for fermions which yields somewhat simpler solutions. For consistency,  the Hamiltonian operator must be Hermitian; hence, the supercharges must satisfy $\bar S=S^\dagger$ and  $S=\bar{S}^\dagger$. The only nonzero (anti)commutators are
\begin{equation}
	[a,\pi_{a}]=[\phi,\pi_{\phi}]=i\hbar,\qquad\{\lambda,\bar{\lambda}\}=\frac{4\pi}{3}l_P^2,\qquad\{\eta,\bar{\eta}\}=-\hbar c,  \label{conmutadores}
\end{equation}
where $l_P^2=\frac{\hbar G}{c^3}$ is the Planck length.  
For the quantization, we redefine the fermionic degrees of freedom as
$\lambda=\sqrt{\frac{\hbar c\kappa^2}{6}}\alpha$, $\bar\lambda=\sqrt{\frac{\hbar c\kappa^2}{6}}\bar\alpha$, $\eta=\sqrt{\hbar c}\beta$
and $\bar\eta=\sqrt{\hbar c}\bar\beta$. Hence, the anticommutators are
\begin{equation}
	\qquad\{\alpha,\bar{\alpha}\}=1,\qquad\{\beta,\bar{\beta}\}=-1.  \label{conmutadoresf}
\end{equation}
as well as $\alpha^{2}=\beta^{2}=\bar{\alpha}^{2}=\bar{\beta}^{2}=0$.
The bosonic momenta are represented by derivatives, $\alpha$ and $\beta$ are 
annihilation operators, and $\bar{\alpha}$ and $\bar{\beta}$ are creation operators.
We fixed the ordering ambiguities by Weyl ordering, which  is antisymmetric for fermions. Hence, the supersymmetric constraint operators read

\begin{align}
	\frac{1}{\sqrt{\hbar c}} S  & =\frac{c\kappa}{2\sqrt{6}}\left(a^{-\frac{1}{2}}\pi_a+\pi_a a^{-\frac{1}{2}}\right)\alpha
		+ca^{-\frac{3}{2}}\pi_\phi\beta
		+\frac{3i\kappa}{\sqrt{6}}  a^\frac{3}{2}W\alpha
		+i a^\frac{3}{2}W'\beta
		-i\frac{\sqrt{6k}}{\kappa}a^\frac{1}{2}\alpha
		-\frac{i\sqrt{3}}{4\sqrt{2}}\hbar c\kappa a^{-\frac{3}{2}}\alpha[\bar{\beta},\beta],\label{Squant}\\
		\frac{1}{\sqrt{\hbar c}} \bar S  & =\frac{c\kappa}{2\sqrt{6}}\left(a^{-\frac{1}{2}}\pi_a+\pi_a a^{-\frac{1}{2}}\right)\bar\alpha
		+ca^{-\frac{3}{2}}\pi_\phi\bar\beta
		-\frac{3i\kappa}{\sqrt{6}}  a^\frac{3}{2}W\bar \alpha
		-i a^\frac{3}{2}W'\bar \beta
		+i\frac{\sqrt{6k}}{\kappa}a^\frac{1}{2}\bar \alpha
		+\frac{i\sqrt{3}}{4\sqrt{2}}\hbar c\kappa a^{-\frac{3}{2}}\bar\alpha[\bar{\beta},\beta].\label{Sbquant}
	\end{align}

Anticommutator  $\{S,\bar S\}=-2\hbar c H_0$ gives the quantum Hamiltonian
	\begingroup\makeatletter\def\f@size{9.5}\check@mathfonts
	\def\maketag@@@#1{\hbox{\m@th\normalsize\normalfont#1}}
	\begin{align}
		H_0=&-\frac{c^2\kappa^2}{24}\left(a^{-1}\pi_a^2+\pi_a^2a^{-1}\right)+\frac{c^2}{2}a^{-3}\pi_\phi^2
		-\frac{\sqrt{3} i}{2\sqrt{2}}\hbar c^2\kappa a^{-3}\pi_\phi(\alpha\bar\beta+\bar\alpha\beta)-\frac{3k}{\kappa^2}a
		-\frac{\sqrt{k}}{4}\hbar ca^{-1}[\alpha,\bar\alpha]+\frac{3\sqrt{k}}{4}\hbar ca^{-1}[\beta,\bar\beta]\nonumber\\
		&-\frac{3\kappa^2}{4}a^3W^2+3\sqrt{k}a^2W+\frac{1}{2}a^3{W'}^2+\frac{3}{8}\hbar c\kappa^2W[\alpha,\bar\alpha]
		-\frac{3}{8}\hbar c\kappa^2W[\beta,\bar\beta]+\frac{\sqrt{3}}{2\sqrt{2}}\hbar c\kappa W'(\alpha\bar\beta-\bar\alpha\beta)
		+\frac{1}{2}\hbar cW''[\beta,\bar\beta]\nonumber\\
		&+\frac{3}{16}(\hbar c\kappa)^2a^{-3}\left(\bar\alpha\alpha\beta\bar\beta+\alpha\bar\alpha\bar\beta\beta\right).\label{Hquant}
	\end{align}
	\endgroup

The Hilbert space is generated from the vacuum state $\ket{1}$, which
satisfies\linebreak \mbox{$\alpha\ket{1}=\beta\ket{1}=0$}. Hence, there are four orthogonal states
\begin{equation}
	\ket{1},\quad\ket{2}=\bar{\alpha}\ket{1},\quad\ket{3}=\bar{\beta}\ket{1}\quad
	\mathrm{and}\quad\ket{4}=\bar{\alpha}\bar{\beta}\ket{1},\label{estados}
\end{equation}
which have norms $\braket{2|2}=\braket{1|1}$, $\braket{3|3}=-\braket{1|1}$ and
$\braket{4|4}=-\braket{1|1}$. Hence, a general state  has  the form
\begin{equation}
	\ket{\psi}=\psi_{1}(a,\phi)\ket{1}+\psi_{2}(a,\phi)\ket{2}+\psi_{3}(a,\phi)\ket{3}+\psi_{4}(a,\phi)\ket{4}.\label{estado}
\end{equation}

Therefore, from  constraint equation $S\ket{\psi}=0$, we obtain
\begin{align}
	a\left(\partial_{a}-\frac{3}{\hbar c} a^{2}W+\frac{6\sqrt{k}}{\hbar c\kappa^2}a+\frac{1}{2}a^{-1}\right)  \psi_{2}
	-\frac{\sqrt{6}}{\kappa}\left(  \partial_{\phi}-a^{3}W'\right)  \psi_{3}=0,\label{ec23}\\
	\left(  \partial_{a}-\frac{3}{\hbar c} a^{2}W+\frac{6\sqrt{k}}{\hbar c\kappa^{2}}a-a^{-1}\right)  \psi_{4}=0\quad \text{and}\quad
	\left(  \partial_{\phi}-\frac{1}{\hbar c}a^{3}W'\right)  \psi_{4}=0,\label{ec4}
\end{align}
while from $\bar{S}\psi=0$
\begin{align}
	a\left(\partial_{a}+\frac{3}{\hbar c} a^{2}W-\frac{6\sqrt{k}}{\hbar c\kappa^2}a+\frac{1}{2}a^{-1}\right)  \psi_{3}
	-\frac{\sqrt{6}}{\kappa}\left(  \partial_{\phi}+a^{3}W'\right)  \psi_{2}=0,\label{ec32}\\
	\left(  \partial_{a}+\frac{3}{\hbar c} a^{2}W-\frac{6\sqrt{k}}{\hbar c\kappa^{2}}a-a^{-1}\right)  \psi_{1}=0,\quad \text{and}\quad
	\left(  \partial_{\phi}+\frac{1}{\hbar c}a^{3}W'\right)  \psi_{1}=0,\label{ec1}
\end{align}

The terms with $a^{-1}$ in (\ref{ec23})--(\ref{ec1}) differ from those  in~\citep{previo:2016} due to a different operator ordering in the Hamiltonian. In fact, the classical Hamiltonian in~\citep{previo:2016} has a term $a^{-1}\pi_a^2$ that can be ordered in many ways to give a Hermitian operator as $a^{-1}\pi_a^2\rightarrow \frac{1}{2k+l}(ka^{-1}\pi_a^2+l\pi_a ^{-1} \pi_a+k\pi_aa^{-1})$.

\subsection{Solutions}\label{sec_sol}
As the Wheeler--DeWitt equation is second-order, its solutions require boundary conditions. However, in  supersymmetric theory, Equations (\ref{ec23})--(\ref{ec1})  are first-order and have unique solutions that can be fixed by consistency and normalization.
The equations for $\psi_{1}$ and $\psi_{4}$ can be straightforwardly solved yielding the unique solutions up to constant factors~\citep{tkachwf}
\begin{align}
	\psi_{1}(a,\phi) &  =A a\exp \left[-\frac{1}{\hbar c} \left( a^{3}W(\phi)-\frac{3\sqrt{k}a^{2}}{\kappa^{2}} \right) \right],\label{psi1aT}\\
	\psi_{4}(a,\phi) &  =A a\exp \left[\frac{1}{\hbar c} \left(  a^{3}W(\phi)-{\frac{3\sqrt{k}a^{2}}{\kappa^{2}}}\right)\right],\label{psi4aT}
\end{align}

The power of  factor $a$ of these solutions differs from the solutions in~\citep{previo:2016} due to the different operator ordering mentioned end of the preceding section. As shown in~\citep{previo:2016}, the solutions of  Equations (\ref{ec23}) and (\ref{ec32}) are not defined at $a=0$ unless they are trivial.
Thus, we chose the solutions
$\ket{\psi}=C_1\,\psi_{1}(a,\phi)\ket{1}+C_4\,\psi_{4}(a,\phi)\ket{4}$,
where $C_1$ and $C_4$ are arbitrary constants ~\citep{previo:2016}.
The norm of this state is
\begin{equation}
	\braket{\psi|\psi}=\left[|C_1|^2\int|\psi_1(a,\phi)|^2dad\phi-|C_4|^2\int|\psi_4(a,\phi)|^2dad\phi\right]\braket{1|1}.\label{norma}
\end{equation}

\begin{figure}[h!]
	\includegraphics[height=4cm,width=6cm]{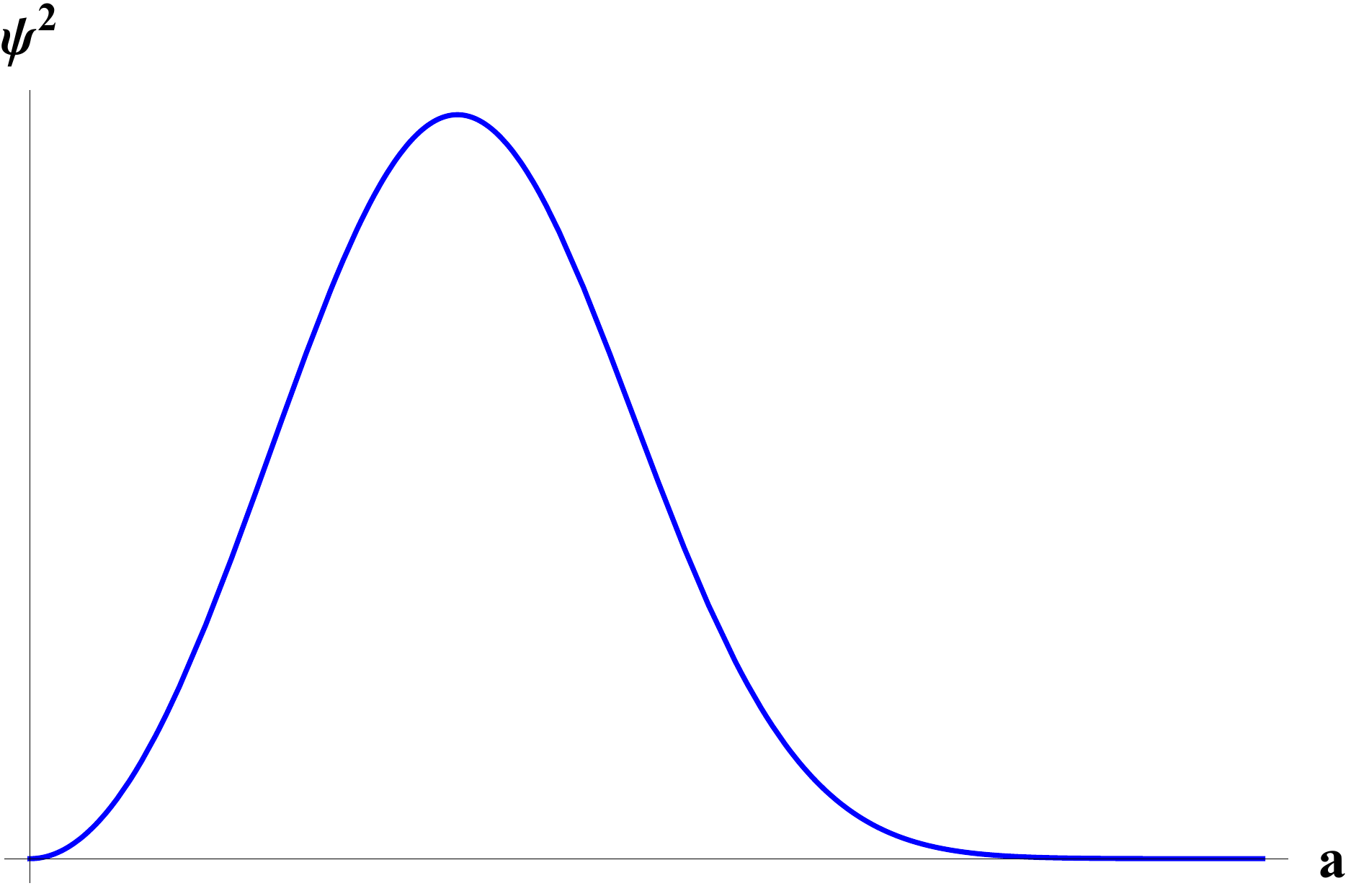}
	\caption{Profile of $\psi^2(a,\phi)$, for $\phi$ constant.}
	\label{fig_campana}
\end{figure}

Classically, $a\geq0$ and could be a problem for quantization, e.g., it could require an infinite wall \cite{Isham}. However, solutions (\ref{psi1aT}) and (\ref{psi4aT}) already vanish at $a=0$. For a positive superpotential, $\psi_{1}$ has a bell form and tends to zero as $a$ increases; see Figure \ref{fig_campana}. In this case,  solution $\psi_{2}$ must be set to be the trivial one. Oppositely, for a negative superpotential, $\psi_{2}$ tends to zero as $a$ increases, and $\psi_{1}$ must be discarded. 
For $\phi\to\pm\infty$, the behavior of  (\ref{psi1aT}) and (\ref{psi4aT}) depends on the form of the superpotential. 
Therefore, we considered only positive or negative superpotentials, and we chose $\braket{1|1}=1$ for positive superpotentials, and $\braket{1|1}=-1$ for negative superpotentials. Hence,
\begin{align}
	\ket{\psi}  &  = C\psi_1(a,\phi)\ket{1},\quad\text{if}\ W(\phi)>0,\label{psi1w}\\
	\ket{\psi}  &  =C\psi_4(a,\phi)\ket{4},\quad\text{if}\ W(\phi)<0. \label{psi4w}
\end{align}

From Expansions (\ref{estado}), and (\ref{estados}), we see that these states correspond to scalars. By construction, these states are invariant under supersymmetry transformations.

Further, for $\phi\to\pm\infty$, the behavior of (\ref{psi1aT}) and (\ref{psi4aT}) depends on the form of the superpotential. For a localized particle, the wave function has well-defined position probabilities and probability conservation. These conditions also guarantee hermicity of operators. On the other side,  the wave functions of free particles do not vanish at infinity, but can be given a meaning by considering relative probabilities. If we restrict the superpotential to be an even function of $\phi$, then operators $\pi_\phi$ and $H_0$ are self-adjoint even if the wave function does not vanish at $\phi\to\pm\infty$. Otherwise, if the wave function vanishes at $\phi\to0$, the domain of $\phi$ can be taken to be $[0,\infty)$. A self-adjoint Hamiltonian constraint is consistent with the lack of evolution in the Heisenberg picture
\begin{equation}
	\bra{\psi}\frac{da}{dt}\ket{\psi}=\frac{i}{\hbar}\bra{\psi}[H_0,a]\ket{\psi}=0.\label{noevol}
\end{equation}

In the following, unless otherwise stated, we consider $k=0$. In this case, we can write either (\ref{psi1w}) or (\ref{psi4w}) as 
\begin{align}
	\psi(a,\phi)  = Ca\exp \left[-\frac{1}{\hbar c} a^{3}|W(\phi)| \right]. \label{psiw}
\end{align}

This wave function differs from the one in~\citep{previo:2016}, by the power of the $a$ in front of the exponential, due to a different operator ordering, as mentioned in the preceding section.

In  Appendix~\ref{appa} we give the expressions for $k=1$.

\section{Time}
\label{time}
Ordinary quantum mechanics assigns operators with real spectra to observables and probability amplitudes for their occurrence. A time dependence of the wave function is generated by the Hamiltonian operator. However, in an actually occurring state, a definite time entails certain energy indeterminacy. Hence, the energy spectrum cannot be restricted to one value, and time-dependent mean values arise by interference among different energy states. In general relativity, the Hamiltonian is constrained to vanish, and time is undetermined. This is consistent, since different energy states can manifest only if there is an environment that makes possible transitions among them. Nonetheless, the universe certainly has different components interacting at least gravitationally, and one such component, suitable chosen, can play the role of a clock~\citep{banks}; see also~\citep{Kuchar,Anderson}. 

In~\citep{previo:2016}, we chose the scalar field as time parameter, and defined an effective time-dependent wave function that allows for computing time-dependent mean values. Here, we analyze  Wave Function (\ref{psiw}) and its probability density regarding the motivation of this choice of time. 

We adopted the standard interpretation of quantum mechanics for the solutions of the Wheeler--DeWitt equation. Hence, the square modulus of the wave function gives the probability density for the possible three-geometries. Further, invariance under time reparametrizations ensures that the superspaces of each of the space slices are equivalent. Thus, this wave function describes, in  a quantum mechanical sense, the space geometries of the whole spacetime. 
Furthermore, if it is possible to identify mean trajectories in superspace along regions around the maxima of the wave function, then it should be possible to parametrize these trajectories following the idea of Misner's supertime~\citep{misner}. These trajectories should be around classical trajectories~\citep{halliwell} corresponding to effective theories. Strictly speaking, measurements should give random values around these mean values. 

In the model of this paper, the configuration space is given by the scale factor and the scalar field. Further, from  (\ref{psiw}), we see that, if we keep $\phi$ constant, the probability density of Wave Function (\ref{psiw}) has the generic bell form of Figure \ref{fig_campana}, with the maximum at
\begin{equation}
	a_{\text{max}}(\phi)=\left[\frac{\hbar c}{3W(\phi)}\right]^{1/3}.\label{amax}
\end{equation}

Allowing  for $\phi$ to vary, these maxima trace out a curve of most probable values of the scale factor. The probability density along this curve is
\begin{equation}
	\psi^2(a_{\text{max}},\phi)=e^{-2/3} a^2_{\text{max}}(\phi),
\end{equation}

Therefore, higher values of the scale factor along this curve have a higher relative probability, and we could think of an expanding universe for this wave function. On the other side, in quantum mechanics mean values are  generally time-dependent, and we can speculate that later times correspond   to higher probabilities here. Hence, time would increase monotonically with the scale factor. Nevertheless, $a_{\text{max}}$ is driven by Scalar Field (\ref{amax}); hence, a natural choice for time is the scalar field, and the superpotential should ensure that $a_{\text{max}}$ increases properly. Such a choice corresponds to a gauge where the scalar field is constant on the spacial slices. In this case, the universe is not localized in the $\phi$ direction and the wave function is not normalizable in this direction, but relative probabilities can be considered, i.e., quotients of probabilities. As the value of $\phi$ is highly uncertain, a measurement gives random values, and we can ask then for the probability that the scale factor takes a value, which is given by the conditional probability of obtaining this value of $a$, given a value of $\phi$ ~\citep{previo:2016}
\begin{equation}
	\left\vert\Psi(a,\phi)\right\vert^2=\frac{\left\vert\psi(a,\phi)\right\vert^2}{\int_0^\infty da\,\vert\psi(a,\phi)\vert^2}.\label{condprob}
\end{equation}

This probability must be used if there is a correlation between both fields, as required for a clock in~\citep{banks}, where it is called relative probability.
Probability (\ref{condprob}) can be obtained~\citep{previo:2016} from a wave function normalized for a given value of $\phi$
\begin{equation}
	\Psi(a,\phi)=\frac{1}{\sqrt{\int_0^\infty da\,\vert\psi(a,\phi)\vert^2}}\,\psi(a,\phi).
	\label{onda}
\end{equation}

Thus, making $\kappa\phi\to t/\mu$  where $\mu$ accounts for the time scale, \protect and taking Solution~(\ref{psiw}), Wave Function (\ref{onda}) becomes
\begin{equation}
	\Psi(a,t)=\sqrt{\frac{6|W(t)|}{\hbar c}}a\exp\left[{-\frac{a^{3}|W(t)|}{\hbar c}}\right],
	\label{ondaw}
\end{equation}
and satisfies a conservation equation~\citep{previo:2016}. Further, as the observed universe is classical, what we can give a meaning to, following the Ehrenfest theorem, is mean values. Thus, under the preceding ansatz, for the scale factor, we obtain~\citep{previo:2016}
\begin{equation}
	a(t)=\int_0^\infty a|\Psi(a,t)|^2 da=\Gamma\left(4/3\right)\left[\frac{\hbar c}{2|W(t)|}\right]^{1/3}, \label{amean}
\end{equation}
which is close to (\ref{amax}), as their quotient is $\Gamma(4/3)(3/2)^{1/3}\sim1.02$. 

The quantum fluctuations produce standard deviations  that satisfy the Heisenberg relation~\citep{previo:2016}.

In this work, we are not producing any approximation, semiclassical or of another type. As the fermionic degrees of freedom do not have classical counterparts, the mean values do not necessarily correspond to trajectories that approximate classical solutions, see e.g.,~\citep{obregon2}.  Actually, (\ref{amean}) can be inserted into the Friedmann equations, from which a potential can be read out. {\protect If we consider the effective FLRW model with a scalar, obtained by inserting~(\ref{amean}) into the Friedmann equations, the corresponding potential can be read out as a time~function}
\begin{equation}
	\frac{1}{3c^4\kappa^2}\left(\frac{2{W'}^{2}}{W^2}-\frac{W''}{W}\right),
\end{equation}
which can be compared with the scalar potential (\ref{scalarpot}),  which follows when the fermionic terms in the Hamiltonian are eliminated.

The results for this section can also be given analytically  for $k=1$; see Appendix~\ref{appa}. They involve hypergeometric, AiryBi, and AiryBi$'$ functions that have exponential behavior, and their numerical evaluation is troublesome. As we were  interested in qualitative features here, we restricted ourselves to $k=0$. In this case, considering that the sign of the superpotential did not have consequences for the wave function or the scalar potential, we chose the superpotential as positive definite in the following.

\section{Inflationary Model}
\label{inflation}
In this section, we consider a class of phenomenological superpotentials that give inflationary dynamics that satisfactorily agrees  with the observational bounds. Potentials of this form appear, e.g., in string-inspired tachyon models. With these superpotentials, the wave function tends to zero as $\phi\to 0$, and we can restrict $\phi\geq0$, as mentioned in Section \ref{sec_sol}. Thus, we consider universes with origin at $t=0$, described by homogeneous quantum cosmology from its very beginning, although above Planck scale, full gravitational interactions become relevant, and some ultraviolet completion would be required. 

In~\citep{previo3}, we performed a qualitative discussion of several superpotentials that have diverse drawbacks, as phantom matter. Here, we consider a family of superpotentials
\begin{equation}\label{superp}
	W(\phi)=\frac{c^4 M_p^3}{\hbar^2}\left[\frac{\sqrt{8\pi}\ \Gamma(4/3)}{2^{1/3}n}\right]^{1/3} \left\{\frac{(\kappa\phi)^{-p}}{\left[1 + e^{(\kappa\phi)^{1/2}}\right]}+ \lambda\right\},
\end{equation}
for $p>0$. The wave function fulfils $\lim_{a\to0}\psi(a,\phi)=0$ and  $\lim_{\phi\to0}\psi(a,\phi)=0$.

Thus, from (\ref{amean}), we have
\begin{equation}\label{at}
	a(t)=n\ell_P\left[\frac{t^p\left(1+e^{\sqrt{t/\mu}}\right)}{\mu^p+\lambda t^p\left(1+e^{\sqrt{t/\mu}}\right)}\right]^{1/3},
\end{equation}
where $\ell_P$ is the Planck length. In the following, we set the normalization constant equal to one, $n=1$, and we redefined the scale factor to be dimensionless, $a/\ell_P\to a$. Constant $\mu$ accounts for the time scale. 
With respect to $\lambda$,  (\ref{amean}) shows  that, if the value of $W$ decreases, the scale factor  grows, and, in order to finish this growth, $W$ must stop  decreasing; this is the role of $\lambda$, which must be positive. Thus, in order to have a sufficiently large increase in  scale factor, $\lambda$ must be small enough. This parameter could be seen as the remainder of a term responsible for dark energy in the superpotential
\begin{equation}
	\frac{\lambda}{1+e^{\beta(\phi-\phi_d)}},
\end{equation}
we do not pursue  this idea further in this work.

For the verification of the predictions of the evolution of Scale Factor (\ref{at}), we consider $N=60$ $e$ folds and observational data of PLANCK observatory~
\begin{itemize}
	\item  Tensor-to-scalar ratio bound $r<0.032$, \citep{tristram},
	
	\item  Scalar spectral index $n_s=0.9649\pm 0.0042$, \citep{planck}.
\end{itemize}

The acceleration, as usual, can be written by  identity
\begin{equation}
	\ddot{a}(t)=a H^2(t)(1-\varepsilon(t)),
\end{equation}
where $\varepsilon(t)=-\frac{\dot{H}(t)}{H^2(t)}$ is the slow-roll parameter. 

The enhancing effect of inflation on inhomogeneous quantum fluctuations is well-known~\citep{baumann}. As a consequence, the fluctuations last after inflation and produce structure seeds, and gravitational waves are generated in the process. These primordial effects last in the CMB, and their study led to stringent bounds on virtual predictions of inflationary models, such as the tensor-to-scalar ratio $r$ and  scalar spectral index $n_s$, that can be evaluated directly from the characteristics of the inflationary evolution.

The tensor-to-scalar ratio is the quotient of the tensor-to-scalar power spectra and is given by
\begin{equation}
	r=16\varepsilon(t).\label{r}
\end{equation}

Scalar spectral index $n_s$ follows from   scalar power spectrum 
\begin{equation}
	{\Delta_s}^2(t)=\left.\frac{1}{8\pi^2{M_P}^2}\frac{H^2(t)}{\varepsilon(t)}\right|_{a(t)H(t)=k},
\end{equation}
as
\begin{equation}\label{ns}
	n_s(t)=1+\left.\frac{d\ln {\Delta_s(t)}^2}{d\ln k}\right|_{a(t)H(t)=k},
\end{equation}
where $k$ are the wavenumbers of the scalar perturbations that, during inflation, exit the horizon at time $t$, i.e., $a(t)H(t)=\dot{a}(t)=k$. Thus, (\ref{ns}) can be written as
\begin{equation}
	n_s(t)=1+\left.k\frac{dt}{dk}\frac{d\ln{\Delta_s(t)}^2}{dt}\right|_{a(t)H(t)=k}=1+\left.\frac{\dot{a}(t)}{\ddot{a}(t)}
	\frac{d\ln{\Delta_s(t)}^2}{dt}\right|_{a(t)H(t)=k}.
\end{equation}
which can also be written  in terms of the Hubble parameter
\begin{equation}
	n_s=\frac{\dot{H}(H^2+5\dot{H})-H\ddot{H}}{\dot{H}(H^2+\dot{H})}.
\end{equation}

Therefore, in order to estimate the values of $r$ and $n_s$ predicted by our model during inflation, we can evaluate them in the dependence of time.

A first approximation for $\lambda$ can be obtained from the limit of very large times of the scale factor, $\lim_{t\to\infty}a(t)=1/\lambda\sim e^N$. $N$ are the $e$ folds generated by inflation for the scale~factor 
\begin{equation}\label{efolds}
	e^N=\frac{a(t_{\text{exit}})}{a(t_{\text{in}})},
\end{equation}
where $t_{\text{in}}$ and $t_{\text{exit}}$ are  the beginning and end times of inflation.
Then, if we compute (\ref{efolds}) with the scale factor (\ref{at}), we can solve for $\lambda$
\begin{equation}\label{lambda1}
	\lambda=\frac{\mu^p}{e^{3 N}-1}\left[\frac{1}{{t_{\text{in}}}^p\left(1+e^{\sqrt{t_{\text{in}}/\mu}}\right)}-\frac{e^{3N}}{{t_{\text{exit}}}^p\left(1+e^{\sqrt{t_{\text{exit}}/\mu}}\right)}
	\right]
\end{equation}

Further, as $\lambda>0$, from (\ref{lambda1}) we obtain a condition for $\mu$. For instance, as $t_{\text{in}}\ll t_{\text{exit}}$, if we set $t_{\text{in}}\ll \mu\ll t_{\text{exit}}$, we can approximate $e^{t_{\text{in}}/\mu}\ll 1$ and $e^{t_{\text{exit}}/\mu}\gg 1$. In this case,  
\begin{equation}\label{mu}
	\mu<\frac{t_{\text{exit}}}{[3N+p(\ln t_{\text{in}}-\ln t_{\text{exit}})]^2}.
\end{equation}

Strictly speaking, we must still define how to define the entry and exit times of inflation depending on the evolution of the scale factor. Relations (\ref{lambda1}) and (\ref{mu}) help in checking the consistency for choices of $\mu$ and $\lambda$. With this in mind, we can fix the values of these parameters and evaluate the behavior in each case. 

In order to get criteria for the entry and exit times, we first analyse the limits of the scale factor, its velocity, and acceleration as $t\to0$.  It turns out as follows. The scale factor tends zero for $p>0$. The velocity tends to $\infty$ for $0<p<3$, for $p=3$ it tends to $\frac{2^{1/3}}{\mu}$, and for $p>3$ it tends to zero. The acceleration tends to $-\infty$ for $0< p<3$, for $3\leq p<6$ it tends to $\infty$, for $p=6$ it tends to $\frac{2^{4/3}}{\mu^2}$, and for $p>6$ it is zero. Additionally, for $t\to\infty$, the scale factor tends to $1/\lambda$, the velocity tends to $0^{+}$, and the acceleration tends to $0^{-}$.

\begin{figure}[h!]
	\includegraphics[width=13cm]{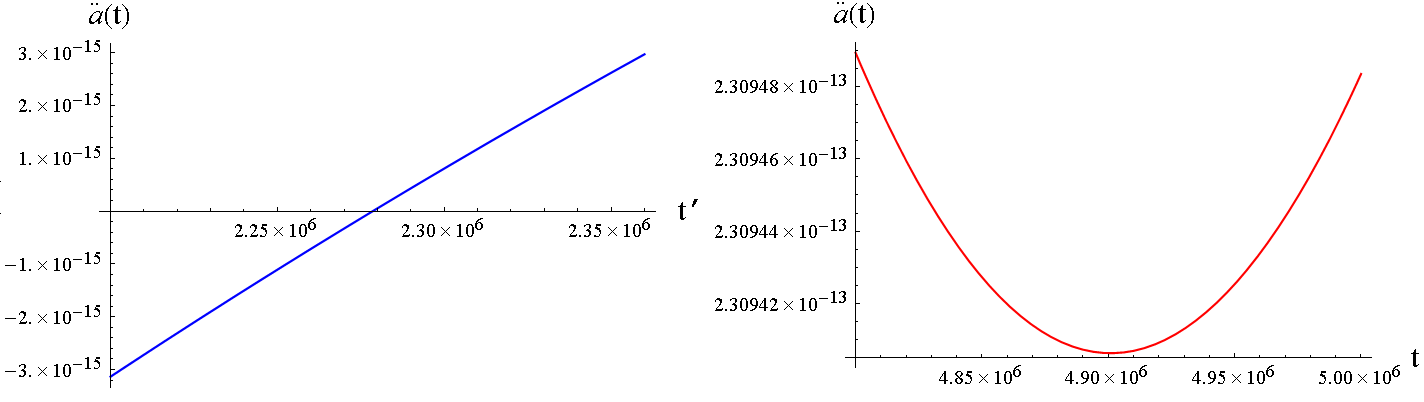}
	\caption{Inflation  entry times for $p=2$ and $p=3$.}
	\label{fig_entry}
\end{figure}

\begin{figure}[h!]
	\includegraphics[width=7cm]{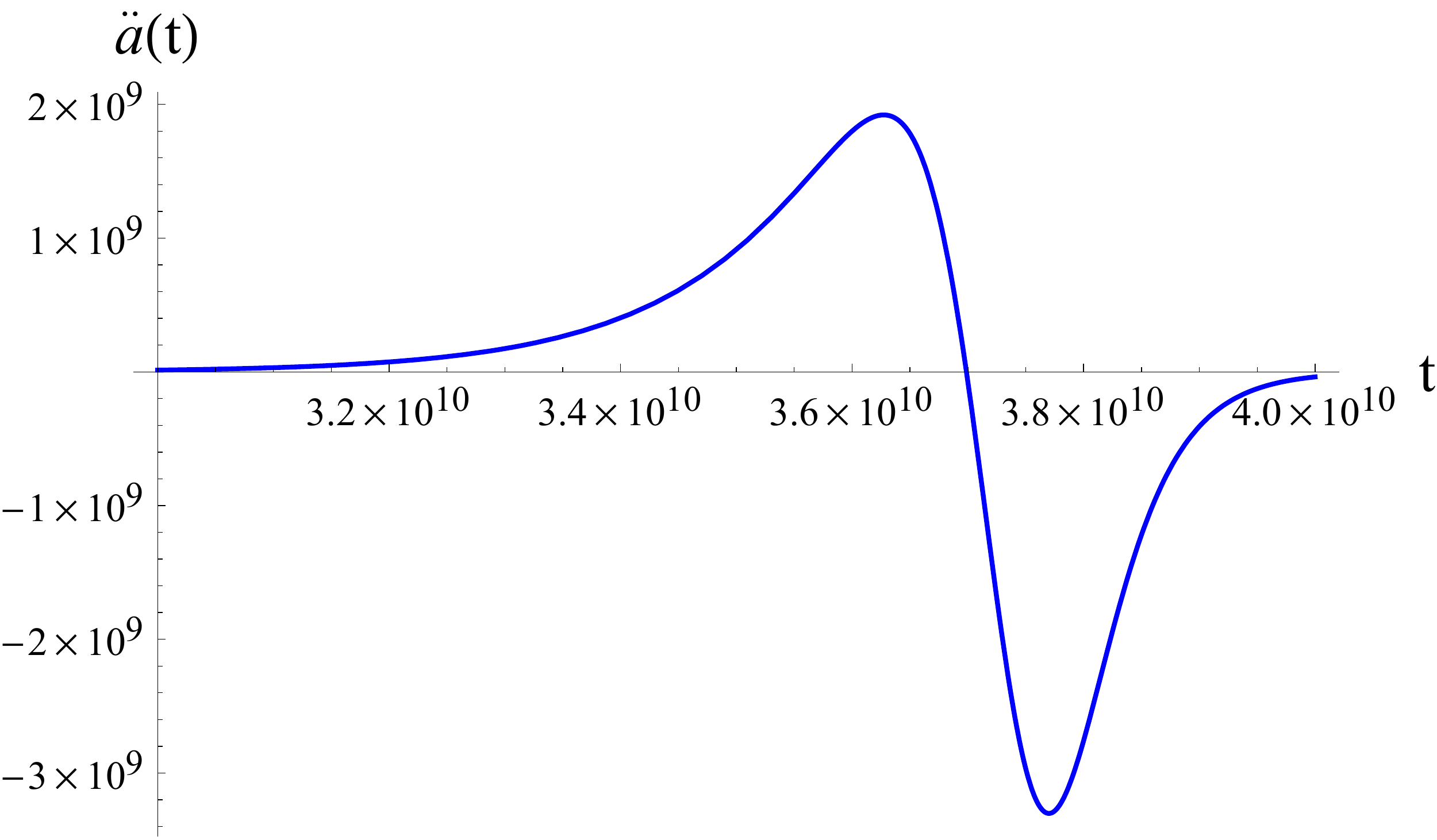}
	\caption{Inflation exit time for $p=1$.}
	\label{fig_exit}
\end{figure}

Considering now the acceleration, if it is negative at $t=0$, for $0<p<3$, it increases until it becomes zero and then positive; we took this zero as the entry time for inflation, see Figure \ref{fig_entry}. Further, the acceleration continues increasing until a maximum later, and after that becomes zero again, and then negative. We took this second zero as the exit time. Afterwards, the acceleration has a minimum, and then tends to zero from the negative numbers; see Figure \ref{fig_exit}. For  cases $3\leq p<6$ where the acceleration tends to infinity at $t=0$, it decreases to a very small minimum that can be taken as the entry time, see Figure \ref{fig_entry}. Further, as in the previous case, it has a maximum and then a zero, and we took it as exit time. For $p\geq 6$, the acceleration at $t=0$ is equal to or larger than zero and increases from the beginning; hence, the entry time is zero, and, as $a(0)=0$, there is no precise way to give an initial time for inflation. For this reason, we considered only $0\leq p<5$.

Further, as a first input, we took the initial time for inflation at a scale lower than the GUT scale  $10^{15}$ GeV. Hence, $t_{\text{in}}\gtrsim 5\times 10^6$~T\textsubscript{P} $\approx 10^{-37}$ s. For the exit of inflation, we took the generally accepted interval $t\sim10^{-33}$--$10^{-32}$ s, and we set  $t_{R}=5\times 10^{-33}$ s = $3.7\times 10^{10}$~T\textsubscript{P}, as reference time for  exit $t_{\text{exit}}\approx t_R$. With these times, for a given value of $p$, we   estimated from (\ref{mu}) and set  $\mu\sim\times 10^6$ T\textsubscript{P}. From it, we computed $\lambda$ from (\ref{lambda1}) for $N=60$, and then evaluated  the entry and exit times from the acceleration $\ddot{a}(t)$. Following this procedure, we adjusted $\mu$ and $\lambda$, so that $t_{\text{exit}}\approx t_R$ and $N=\ln a(t_{\text{exit}})/a(t_{\text{in}})\approx 60$. With these data, we computed the tensor to scalar ratio (\ref{r}), and the scalar spectral index (\ref{ns}). We performed these steps for $p=1,2,3,4,5$. 

It is remarkable that, with the parameters fixed in the previous way, for given \mbox{$0<p\leq 5$}, the entry time is of the order of $10^6$ T\textsubscript{P}, and   the tensor/scalar ratio and the scalar spectral index agree quite well with the observational constrains. 

\begin{figure}[h!]
	\includegraphics[width=7cm]{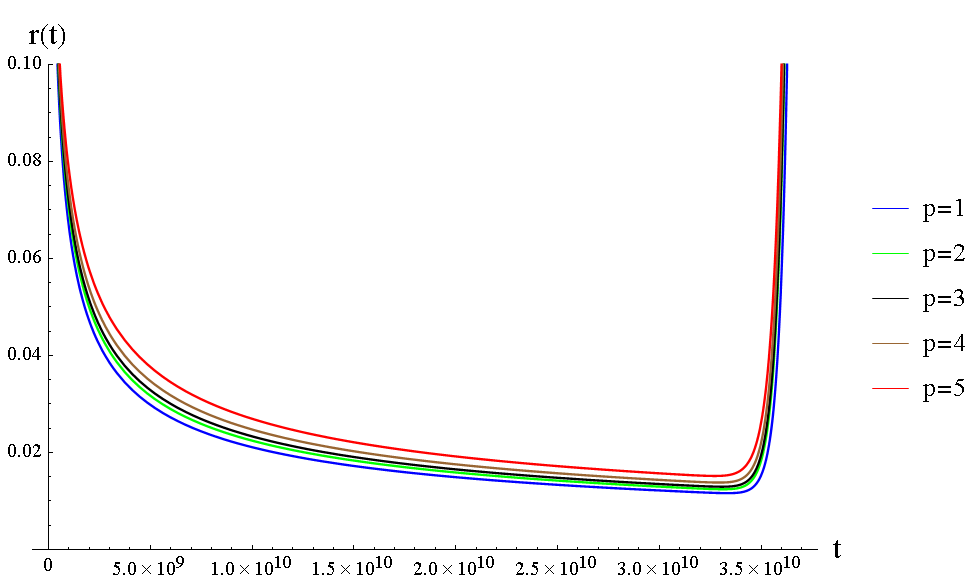}
	\caption{Tensor/scalar ratio.}
	\label{fig_r}
\end{figure}

Tensor/scalar Ratio (\ref{r}) (see Figure \ref{fig_r}) satisfies    bound $r<0.032$ \citep{tristram} very well and depends slightly on $p$. 

\noindent

\begin{figure}[h!]
	\includegraphics[width=7cm]{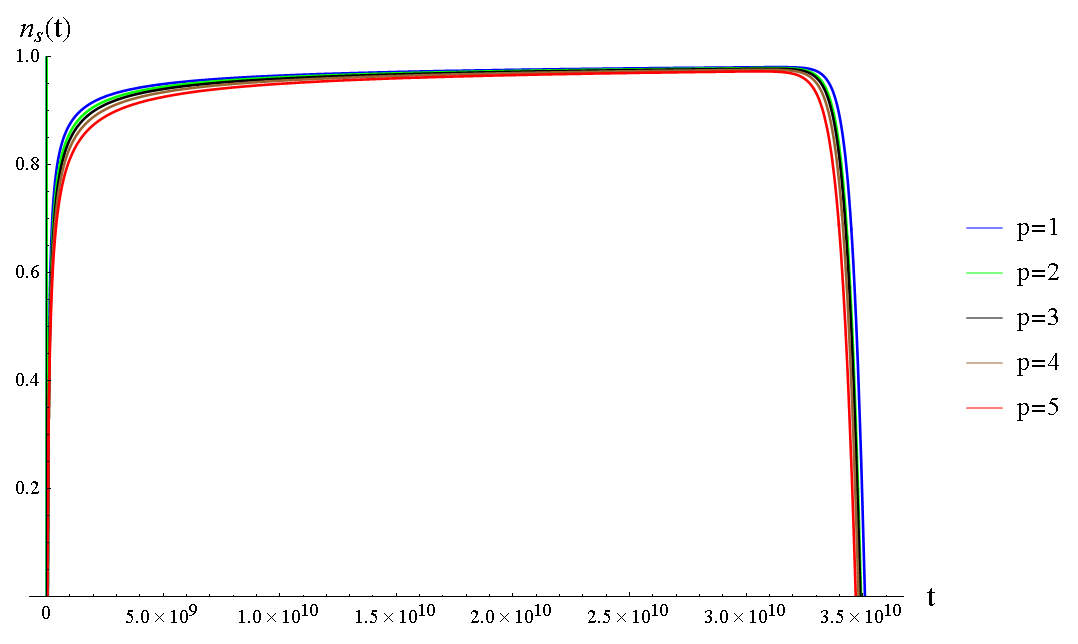}
	\caption{Scalar spectral index.}
	\label{fig_ns}
\end{figure}

Further, from (\ref{ns}), we computed the scalar spectral index whose behavior is shown in Figure \ref{fig_ns}. 

\begin{figure}[h!]
	\includegraphics[width=7cm]{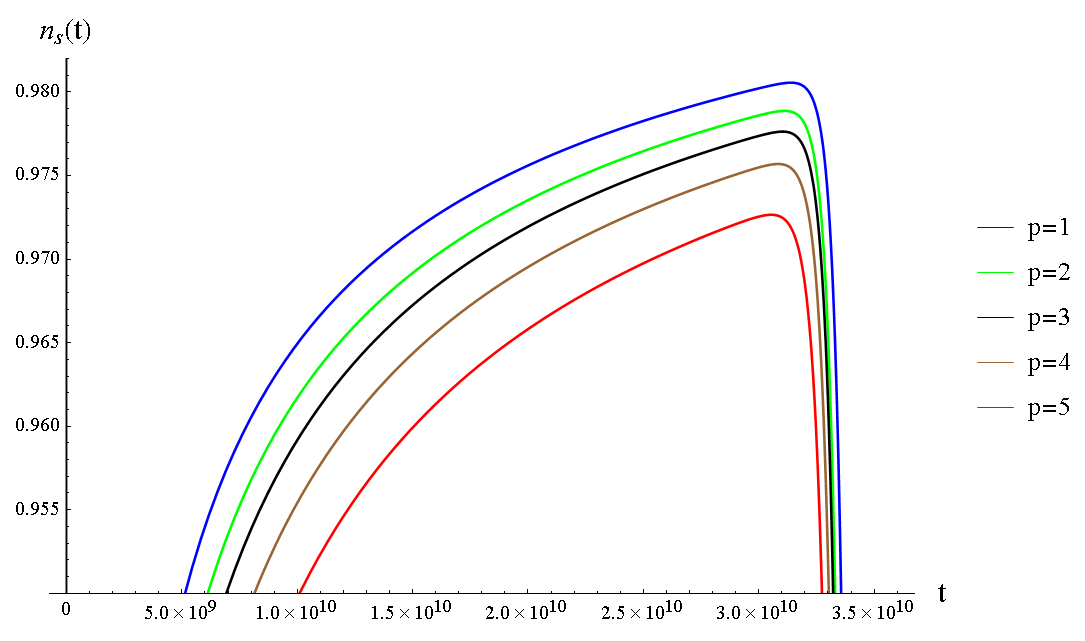}
	\caption{Detail of scalar spectral index.}
	\label{fig_ns2}
\end{figure}

Here, the agreement with  observational bound $0.9649 \pm 0.0042$ (with $68\%$ CL)~\citep{planck}, was very good for most of the duration of inflation, as shown in Figure \ref{fig_ns2}. It is also  slightly dependent of $p$. Note that the running of the scalar spectral index $\frac{d n_s}{d \ln k}$ is positive, with a value around $0.0004$, with a a negative running of running.
\section{Discussion}
\label{discussion}

\textls[-25]{Quantum cosmology gives a canonical quantization of general relativity in the Schr\"odinger picture, }with the Wheeler--DeWitt equation as  a time-independent Schr\"odinger equation. The wave function gives the probability amplitudes for the occurrence of all possible spatial three-geometries and field distributions, at any time. On the other hand, the observed universe is classical and has time, and classical physics follows from quantum physics, hence an effective time evolution should follow from the quantum description. Further, the Wheeler--DeWitt equation is a second-order scalar equation. In the supersymmetric case, there is a system of first-order equations whose solutions are spinorial wave functions, see, e.g.,~\citep{eath,moniz-1}. A particular meaning of the components of these wave functions has not been given; it would quickly lead to a Machian discussion. In many cases, there are only two nonvanishing components, see, e.g.,~\citep{obregon1}, given by real exponentials with opposite exponent sign. 
Here, we considered an FLRW theory with a minimally coupled scalar field. Single-field models are extremely effective to account for the inflationary~era.

From the four components of the solution of the supersymmetric Wheeler--DeWitt equation, only one tends to zero as $a\to\infty$, and the other can be taken to be trivial. The form of Solution (\ref{psiw}) suggests the ansatz that, in a certain gauge, time can be given by the scalar field, and trajectories for the observables are mean values on an effective wave function that corresponds to conditional probabilities. Thus, for the scale factor, we obtain an evolution $a(t)$, and we can perform time reparametrizations considering that it is scalar $a'(t')=a(t)$. The resulting evolution is nonperturbative. These trajectories are classical if the quantum fluctuations are negligible. The time proposed here does not follow from a semiclassical approach. 

We  considered in this formulation {\protect one  family of inflationary models that depends on the parameter $p>0$; as for $p>5$ there is eternal inflation, we restrict it to $p<5$ and for simplicity consider only integer values. There are other two parameters,  $\mu$ and $\lambda$; the first  fixes the time scale. Parameter $\lambda$ corresponds to the extent of the inflation, the $e$ folds. We convened to set the entry of inflation, depending on the value of $p$, at zeros or minima of the acceleration, and exit times as zeros. In this way, for all $p$, we adjusted parameters $\mu$ and $\lambda$, so that the exit time coincided with a reference time $t_R=5\times 10^{-33} s$, and $N\approx 60$. With this setting, the corresponding tensor-to-scalar ratio and scalar spectral index can be computed with values that agree with the observational bounds remarkably well}.

For indepth analysis, inhomogeneous perturbations should be introduced to evaluate their evolution and the the production of primordial gravitational waves. These computations should be performed at best in the superfield formalism. An interesting perspective is also the study of this formalism for dark energy. 

\vskip 1truecm

\centerline{\bf Acknowledgements}
C.R. thanks O. Obreg\'on and H. Garc\'{\i}a-Compe\'an, and  V. V.  thanks G. Garc\'{\i}a-Arroyo and J.A. V\'azquez
for the interesting discussions. This research was funded by VIEP-BUAP. N.E. Martínez-
Pérez thanks CONACyT for studies grant.
conflict of interest.

\appendix

\section[\appendixname~\thesection]{}\label{appa}
In this appendix, we give,  for $k=1$, the expressions for the normalization factor for the wave function (\ref{psi1aT}) for $k=1$, and the time-dependent scale factor (\ref{amean}).
The normalization factor of
\begin{equation}
	\psi(a,\phi)=|C|a\exp\left[\frac{1}{\hbar c}\left(  -a^{3}|W(\phi)|+\frac{3\sqrt{k}a^{2}}{\kappa^{2}}\right)\right],\label{fonda1}
\end{equation}
is given by
\begin{align}
	|C|^{-2}=&\frac{c\hbar}{6} \int_{-\infty}^{\infty}\frac{1}{|W(\phi)|}\Bigg[\, _2F_2\left(\frac{1}{2},1;\frac{1}{3},\frac{2}{3};\frac{8}{c\hbar\kappa^6 W^2(\phi)}\right)
	+4\pi\left[\frac{2}{9c \hbar \kappa ^6W(\phi)^2}\right]^{1/3}e^{\frac{4}{c\hbar\kappa^6 W^{2}(\phi)}}\nonumber\\
	&\times\bigg[\text{Bi}'\left(\left[\frac{6}{c \hbar \kappa ^6W(\phi)^2}\right]^{2/3}\right)
	+\left[\frac{6}{c \hbar \kappa ^6W(\phi)^2}\right]^{1/3}\text{Bi}\left(\left[\frac{6}{c \hbar \kappa ^6W(\phi)^2}\right]^{2/3}\right)\bigg]\Bigg]d\phi, \label{C_1}
\end{align}
where $Bi$ is the Airy function of second kind. 

For the denominator of (\ref{onda})
\begin{align*}
	\int_0^\infty da\,\vert\psi(a,\phi)\vert^2=&\frac{c \hbar }{18 W(\phi )} 
	\times\Bigg\{3\text{  }\, _2F_2\left(\frac{1}{2},1;\frac{1}{3},\frac{2}{3};-\frac{8}{c\hbar \kappa ^6   W(\phi)^2}\right)
	+4\times 6^{1/3} \pi  \left[\frac{1}{c \hbar \kappa ^6W(\phi )^2}\right]^{2/3}e^{-\frac{4}{c \hbar  \kappa ^6 W(\phi )^2}}\\
	&\times\Bigg[6^{1/3} \text{Bi}\left(\left[\frac{6}{c\hbar \kappa ^6W(\phi )^2}\right]^{2/3}\right)
	- \left[c \hbar \kappa ^6W(\phi )^2\right]^{1/3} \text{Bi}'\left(\left[\frac{6}{c \hbar \kappa ^6W(\phi)^2}\right]^{2/3}\right)\Bigg]\Bigg\},
\end{align*}
from which follows
\begin{align*}
	a(t)=\Bigg\{9 \sqrt[3]{3} \kappa ^4 (c \hbar )^{2/3} W(t)^{4/3} \, _2F_2\left(1,\frac{3}{2};\frac{2}{3},\frac{4}{3};\frac{8}{c \kappa^6 \hbar  W(t)^2}\right)
	e^{-\frac{4}{c\hbar  \kappa ^6  W(t)^2}}
	+2^{2/3} \pi \text{  }\left[24+c \kappa ^6 \hbar  W(t)^2\right] \text{Bi}\left(\left[\frac{6}{c \hbar \kappa ^6W(t)^2}\right]^{2/3}\right)\\
	+8 \sqrt[3]{2} 3^{2/3} \pi  \kappa ^2 \sqrt[3]{c \hbar } W(t)^{2/3}\text{  }\text{Bi}'\left(\left[\frac{6}{c \hbar \kappa ^6W(t)^2}\right]^{2/3}\right)\Bigg\}/
	\Bigg\{3 \sqrt[3]{3} \kappa ^6 (c \hbar )^{2/3} W(t)^{7/3} \, _2F_2\left(\frac{1}{2},1;\frac{1}{3},\frac{2}{3};\frac{8}{c \kappa ^6 \hbar  W(t)^2}\right)
	e^{-\frac{4}{c\kappa ^6 \hbar  W(t)^2}}\\
	+4 \sqrt[3]{2} 3^{2/3} \pi  \kappa ^4 \sqrt[3]{c \hbar } W(t)^{5/3}\text{  }\text{Bi}'\left(\left[\frac{6}{c \hbar \kappa ^6W(t)^2}\right]^{2/3}\right)
	+12\ 2^{2/3} \pi  \kappa ^2 W(t)\text{  }\text{Bi}\left(\left[\frac{6}{c \hbar \kappa ^6W(t)^2}\right]^{2/3}\right)\Bigg\}.
\end{align*}

Due to the exponential behavior of the hypergeometric and Airy functions that appear in the numerator and denominator in these expressions, it is difficult to handle them numerically.

\end{document}